\documentclass[preprint,showpacs,amsmath,amssymb]{revtex4}

\usepackage{graphicx}
\usepackage{graphics}

\begin{document}


\title{Analysis of a long-range random field quantum
antiferromagnetic Ising model}
\author{$^{1,2}$Bikas K. Chakrabarti}
\email[e-mail: ]{bikask.chakrabarti@saha.ac.in}
\author{$^{1}$Arnab Das}
\email[e-mail: ]{arnab.das@saha.ac.in}
\author{$^{2}$Jun-ichi Inoue}
\email[e-mail: ]{j_inoue@complex.eng.hokudai.ac.jp}
\affiliation{$^{1}$Theoretical Condensed Matter Physics Division, 
Saha Institute of Nuclear Physics, 1/AF Bidhannagar,
Kolkata-700064, India \\
$^{2}$Complex Systems Engineering, Graduate School of Information
Science and Technology, Hokkaido University, 
N14-W9, Kita-ku, Sapporo 060-0814, Japan}

\begin{abstract}
We introduce a solvable quantum 
antiferromagnetic model. 
The model, with Ising spins 
in a transverse field, 
has infinite range 
antiferromagnetic interactions 
and random fields on each site
following an arbitrary 
distribution. 
As is well-known, 
frustration in the random field Ising model 
gives rise to a many valley structure in the 
spin-configuration space. 
In addition, the antiferromagnetism also
induces a regular frustration even for 
the ground state. 
In this paper, 
we investigate analytically the critical phenomena in the model, 
having both randomness and frustration
and we report some 
analytical results for it.
\end{abstract}


\maketitle

\section{Introduction}
With the realization in the mid last century that the 
N\'{e}el state cannot be the ground state (not even an eigenstate)
of a quantum Heisenberg antiferromagnet, considerable effort has
gone in search of, and in understanding the nature of, the ground state
of such and similar quantum antiferromagnet \cite{Frad}.  
Since early 1960s, 
quantum spin systems 
described by Ising model 
in a transverse tunneling field was investigated extensively;
particularly because of easy mapping of the quantum system
to its equivalent classical system and some cases of
exact solubility \cite{Bikas}. 
However there have, so far, been very few soluble models with 
antiferromagnetic interactions. 
It is well-known that the Ising model 
with long range interactions is 
solved exactly, even if the system has some special kind of
quenched disorder, like in Sherrington-Kirkpatrick model of 
spin glasses. 
The number of degenerate states there 
can be estimated to be 
${\cal O}(2^{N/2})$, which is larger than 
that of the above mentioned model 
(${\cal O}(2^{0.28743N})$). 
However, 
it is not so easy to 
consider the antiferromagnetic version of 
the model due to a lack of 
sub-lattice to capture the N\'{e}el ordering 
at low temperatures. 
In this paper, we introduce and study a solvable 
quantum antiferromagnetic model. 
In our model system  
each spin is influenced 
by the infinite range antiferromagnetic 
interactions in a transverse field. 
We also consider the case under 
the Gaussian or the binary random fields.
By introducing 
two sub-groups of the spin system, 
we describe the system by means of the effective 
single spin Hamiltonian 
which is derived by the Trotter decomposition \cite{Suzuki} and 
Hubberd-Stratonovich transformation \cite{Chaikin}, and
we solve the model analytically. 
A preliminary study, for the case without
random fields, was reported earlier \cite{Cond}.

This paper is organized as follows. 
In the next section, we 
introduce our model system and 
write down the general formula of the 
averaged free energy density. 
In section 3, to check 
the validity of our analysis, 
we compare our result with the 
previous well-known result which was 
obtained by mean-field approximations \cite{Kittel}. 
In section 4, 
we consider the system under the Gaussian 
and the binary random fields on site and 
derive the equations of 
states and evaluate them. 
We then obtain the phase diagrams. 
Section 5 gives a summary.
\section{The model system and its analysis}
In order to capture the N$\acute{\rm e}$el ordering  
below the critical 
temperature $T_{N}$ or 
the amplitude of 
transverse field $\Gamma_{N}$, 
we divide spins 
$\mbox{\boldmath $S$}$ into 
the sub group $A$ : 
$\mbox{\boldmath $S$}^{(A)}=
(S_{1}^{x,z,(A)},
S_{2}^{x,z,(A)},
\cdots, 
S_{N}^{x,z,(A)})$ and 
the sub group $B$ : 
$\mbox{\boldmath $S$}^{(B)}=
(S_{1}^{x,z,(B)},
S_{2}^{x,z,(B)},
\cdots, 
S_{N }^{x,z,(B)})$, which are 
corresponding to 
{\it virtual sub-lattice} $A$ and B. 
Then the system 
is described by the following 
effective Hamiltonian  
\begin{equation}
H(\mbox{\boldmath $\tau$})  =  
\frac{J}{N}
\sum_{ij}
S_{i}^{z,(A)}
S_{j}^{z,(B)}
-\sum_{l=A,B} 
\Gamma_{l}\sum_{i}
S_{i}^{x,(l)}
-h\sum_{l=A,B}\sum_{i} h_{i}S_{i}^{z,(l)},
\label{eq:Hamiltonian}
\end{equation}
where $S_{i}^{z,(A,B)},
S_{i}^{x,(A,B)}$ are 
$x$ and $z$ components of 
the Pauli matrix : 
\begin{eqnarray}
S_{i}^{x,(A,B)} =
\left(
\begin{array}{cc}
0  &  1 \\
1  &  0
\end{array}
\right), \,\,\,
S_{i}^{z,(A,B)} = 
\left(
\begin{array}{cc}
1  &  0 \\
0  & -1
\end{array}
\right),
\end{eqnarray}
and $\mbox{\boldmath $h$}=(h_{1},h_{2},
\cdots,h_{N})$ is 
a vector of the 
random fields on site and 
$h$ means the strength of the random field. 
$\Gamma_{A}$ and 
$\Gamma_{B}$ are amplitudes of
the transverse fields in the 
sub groups $A$ and B. 
The main advantage of this Hamiltonian is that it can be
recast exactly to that of a single spin in an effective
field.

Using the Suzuki-Trotter formalism \cite{Suzuki} 
one can express the quenched-variable $\mbox{\boldmath $h$}$-dependent
partition function as
\begin{eqnarray}
Z(\mbox{\boldmath $h$}) & = & 
\lim_{M \to \infty}
{\rm tr}_{\mbox{\boldmath $S$}_{A},
\mbox{\boldmath $S$}_{B}}
{\rm e}^{-\beta H(\mbox{\boldmath $h$})},
\end{eqnarray}
with
\begin{eqnarray}
\beta H(\mbox{\boldmath $h$}) & = & 
\frac{\beta J}{2NM}
\sum_{k}
\left\{
\sum_{i}S_{ik}^{(A)}
+
\sum_{i}S_{ik}^{(B)}
\right\}^{2}-
\frac{\beta J}{2NM}
\sum_{k}
\left\{
\sum_{i}S_{ik}^{(A)}
-
\sum_{i}S_{ik}^{(B)}
\right\}^{2} \nonumber \\
\mbox{} & - & 
\frac{\beta h}{M}\sum_{i,k}
h_{i}S_{ik}^{(A)}
-
\frac{\beta h}{M}
\sum_{i,k}
h_{i}S_{ik}^{(B)}
-
\gamma_{A}\sum_{i,k}S_{ik}^{(A)}S_{ik+1}^{(A)}
-
\gamma_{B}\sum_{i,k}S_{ik}^{(B)}S_{ik+1}^{(B)},
\end{eqnarray}
\noindent where $\gamma_{A,B} =
\frac{1}{2}
\log \coth
\left(
\frac{\beta \Gamma_{A,B}}{M}
\right).$ 
By using 
the Hubberd-Stratonovich 
transformation \cite{Chaikin}, 
the field $\mbox{\boldmath $h$}$-dependent 
partition function is written 
by means of the saddle point technique, 
in the limit $N \to \infty$, as 
\begin{eqnarray}
Z(\mbox{\boldmath $h$}) & = & 
{\rm tr}_{
\mbox{\boldmath $S$}_{A},
\mbox{\boldmath $S$}_{B}}
\int_{-\infty}^{\infty}
\int_{-i\infty}^{+i\infty}
\prod_{k=1}^{M}
\frac{idm_{+}^{k}dm_{-}^{k}}
{(2\pi/N\beta J)}\,
{\rm e}^{\frac{N\beta J}{2M}\sum_{k}(m_{+}^{k})^{2}
-\frac{N\beta J}{2M}\sum_{k}(m_{-}^{k})^{2}} \nonumber \\
\mbox{} & \times & 
{\exp}
{\Biggr [}
\frac{\beta J}{M} \sum_{i,k} m_{+}^{k}
(S_{ik}^{(A)}+
S_{ik}^{(B)})
+
\frac{\beta J}{M} 
\sum_{i,k}
m_{-}
(S_{ik}^{(A)}-
S_{ik}^{(B)}
) \nonumber \\
\mbox{} & + & 
\frac{\beta h}{M} 
\sum_{i,k}\tau_{i}
S_{ik}^{(A)}
+
\frac{\beta h}{M} 
\sum_{i,k} h_{i}
S_{i,k}^{(B)}
+
\gamma_{A}
\sum_{i,k}S_{ik}^{(A)}S_{ik+1}^{(A)}
+
\gamma_{B}
\sum_{i,k}S_{ik}^{(B)}S_{ik+1}^{(B)}
{\Biggr ]} \nonumber \\
\mbox{} & \simeq & 
{\rm e}^{\frac{N\beta J}{2M}\sum_{k} (m_{+}^{k})^{2}-
\frac{N \beta J}{2M} (m_{-}^{k})^{2}} \nonumber \\
\mbox{} & \times & 
{\rm tr}_{
\mbox{\boldmath $S$}_{A},
\mbox{\boldmath $S$}_{B}}
{\exp}
{\Biggr [}
\frac{\beta J}{M} \sum_{i,k} m_{+}^{k}
(S_{ik}^{(A)}+
S_{ik}^{(B)})
+
\frac{\beta J}{M} \sum_{i,k} m_{-}^{k}
(S_{ik}^{(A)}-
S_{ik}^{(B)}
) \nonumber \\
\mbox{} & + & 
\frac{\beta h}{M} 
\sum_{i,k} h_{i}
S_{ik}^{(A)}
+
\frac{\beta h}{M} 
\sum_{i,k} h_{i}
S_{ik}^{(B)}
+
\gamma_{A}
\sum_{i}S_{ik}^{(A)}S_{ik+1}^{(A)}
+
\gamma_{B}
\sum_{i,k}S_{ik}^{(B)}S_{ik+1}^{(B)}
{\Biggr ]} \nonumber \\
\mbox{} & = & 
{\rm e}^{\frac{N\beta J}{2M}\sum_{k} (m_{+}^{k})^{2}-
\frac{N \beta J}{2M} (m_{-}^{k})^{2}} 
{\rm tr}_{
\mbox{\boldmath $S$}_{A},
\mbox{\boldmath $S$}_{B}}
\prod_{i=1}^{N}
{\exp}
\left[
-\beta H_{i}(h_{i})
\right],
\end{eqnarray}
where $H_{i}(h_{i})$ is 
the effective single spin Hamiltonian 
and is given by 
\begin{eqnarray}
H_{i}(h_{i}) & = & 
-\frac{J}{M}
\sum_{k}
m_{+}^{k}
(S_{ik}^{(A)}+S_{ik}^{(B)})
-
\frac{J}{M}
\sum_{k} m_{-}^{k}
(S_{ik}^{(A)}-S_{ik}^{(B)}) \nonumber \\
\mbox{} & - & 
\frac{hh_{i}}{M}
\sum_{l=A,B}\sum_{k}S_{ik}^{(l)}
-\beta^{-1}
\sum_{l=A,B} \gamma_{l}
S_{ik}^{(l)}
S_{ik+1}^{(l)}.
\label{eq:Single_Ham}
\end{eqnarray}
Here we should keep in mind that 
the integrals with respect to 
$m_{+}^{k}$ and $m_{-}^{k}$ are 
evaluated at the saddle points 
in the limit of $N \to \infty$, 
namely  
\begin{eqnarray}
m_{+}^{k} & = & 
-\frac{1}{N}
\sum_{i}S_{ik}^{(A)}
+
\frac{1}{N}
\sum_{i}S_{ik}^{(B)}=
-(M_{A}^{k,z}+M_{B}^{k,z}) 
\label{eq:m_to_M1} \\
m_{-}^{k} & = & 
\frac{1}{N}
\sum_{i}S_{ik}^{(A)}
-
\frac{1}{N}
\sum_{i}S_{ik}^{(B)} = (M_{A}^{k,z}-
M_{B}^{k,z}),
\label{eq:m_to_M2}
\end{eqnarray}
where $m_{+}^{k}$ and $m_{-}^{k}$ are related to 
the magnetizations of 
$z$-component for 
group $A$ and $B$, namely 
$M_{A}^{k,z}$ and 
$M_{B}^{k,z}$.
Thus the free energy 
$F(\mbox{\boldmath $h$})=-\beta^{-1}
\log Z(\mbox{\boldmath $h$})$ 
of the system is now 
written by 
\begin{equation}
F(\mbox{\boldmath $h$}) = 
-\frac{N J}{2M}\sum_{k}(m_{+}^{k})^{2} +
\frac{N J}{2M}\sum_{k}(m_{-}^{k})^{2} -  
\beta^{-1} \log 
\prod_{i=1}^{N}
{\rm tr}_{S_{ik}^{(A)},
S_{ik}^{(B)}}
{\exp}
\left[
-\beta H_{i}(h_{i})
\right].
\end{equation}
Taking into account the symmetry of the system, 
$m_{+}^{k}=m_{+}, 
m_{-}^{k}=m_{-}$ for all $k$, i.e.,
the so-called static approximation holds good 
naturally; the fluctuations due to, say,
two-spin correlations (including entanglements) vanishes 
in thermodynamic limit. 
Thus 
the $\mbox{\boldmath $h$}$-dependent free energy leads to 
\begin{equation}
F(\mbox{\boldmath $h$})  =  
-\frac{N J}{2}(M_{A}^{z}+M_{B}^{z})^{2} + 
\frac{N J}{2}(M_{A}^{z}-M_{B}^{z})^{2}  
-\beta^{-1} \log \prod_{i=1}^{N}
{\cal Z}_{i}^{(A)}
{\cal Z}_{i}^{(B)}
\end{equation}
with 
\begin{equation}
{\cal Z}_{i}^{(A,B)} = 
{\rm tr}_{S_{i}^{x,z,(A,B)}}
{\rm e}^{\beta (
-2JM_{B,A}^{z}+h h_{i})S_{i}^{z,(A,B)}
+\beta \Gamma_{A,B}S_{i}^{x,(A,B)}},
\label{eq:MtoInf}
\end{equation} 
where we used the relations 
(\ref{eq:m_to_M1}) and (\ref{eq:m_to_M2}): 
$m_{+}+m_{-}=-2M_{B}^{z}, m_{+}-m_{-}=-2M_{A}^{z}$. 
It should be noted that 
the above free energy still depends on the 
fields $\mbox{\boldmath $h$}$. 
To obtain the $\mbox{\boldmath $h$}$-independent 
averaged free energy $F$, we should 
evaluate the following quantity 
\begin{eqnarray}
F & = & 
\int_{-\infty}^{\infty}
P(\mbox{\boldmath $h$})
F(\mbox{\boldmath $h$})
d \mbox{\boldmath $h$},
\label{eq:ave_Free} 
\end{eqnarray}
where 
$P(\mbox{\boldmath $h$}) = 
P(h_{1},\cdots,h_{N})$ is a joint distribution of 
the random fields and we defined 
$d \mbox{\boldmath $h$}=
dh_{1} \cdots dh_{N}$. 
If we assume that the random field $h_{i}$ for 
each site $i$ is uncorrelated (not influenced by other 
$h_{j}{\rm{'s}}; ~ j\neq i$) then
\begin{eqnarray}
P(\mbox{\boldmath $h$}) & = & 
P(h_{1},\cdots,h_{N}) = 
P(h_{1})\cdots P(h_{N})=
\prod_{i}P(h_{i}),
\end{eqnarray}
there in the average in (\ref{eq:ave_Free}), 
giving the free energy (per spin)  
\begin{equation}
f = \sum_{l=A,B}f_{l},\,\,
f_{l} = 
- J M_{l}^{z}M_{l+1}^{z} -\beta^{-1}
\int_{-\infty}^{\infty}
d\hat{h} P(\hat{h}) 
\log 2 \cosh 
\beta \sqrt{(2JM_{l}^{z}-h\hat{h})^{2}+\Gamma_{l+1}^{2}}.
\label{eq:free_dens}
\end{equation}
Here we used the fact that 
the $2 \times 2$ matrix 
$\mbox{\boldmath $H$}$ (with 
the elements 
$(\mbox{\boldmath $H$})_{11}=
-(\mbox{\boldmath $H$})_{22}=a, 
(\mbox{\boldmath $H$})_{12}=
(\mbox{\boldmath $H$})_{21}=b$ 
appearing in the of exponent of 
equation (\ref{eq:MtoInf})) 
has 
eigen values $\pm \sqrt{a^{2}+b^{2}}$. 
We also should keep in mind that 
in the sum with respect to $l$, 
$A+1=B,B+1=A$; hence $|A-B|=1$ is satisfied. 
Hereafter, we use this relation 
for the sum with respect to 
the label $l$. 
The magnetizations for two sub-lattices 
$M_{A}^{z}$ and $M_{B}^{z}$ now obey the 
following saddle point equations  
\begin{equation}
M_{l} =  
\int_{-\infty}^{\infty}
d\hat{h} P(\hat{h}) 
\frac{-(2JM_{l+1}^{z}-h\hat{h})}
{\sqrt{(2JM_{l+1}^{z}-h\hat{h})^{2}+
\Gamma_{l}^{2}}}
\tanh 
\beta 
\sqrt{
(2JM_{l+1}^{z}-h\hat{h})^{2}+
\Gamma_{l}^{2}},\,
(l=A,B)
\label{eq:Ma} 
\end{equation}
where it should be noted that 
$A-B=1$ and the appropriate expression for magnetization
of each sub-lattice can be derived by
equating  
$\partial f/\partial M_{B}^{z}$ and 
$\partial f/\partial M_{A}^{z}$ to zero 
respectively. 
\section{Analysis under uniform field}
\begin{figure}[ht]
\begin{center}
\includegraphics[width=8.6cm]{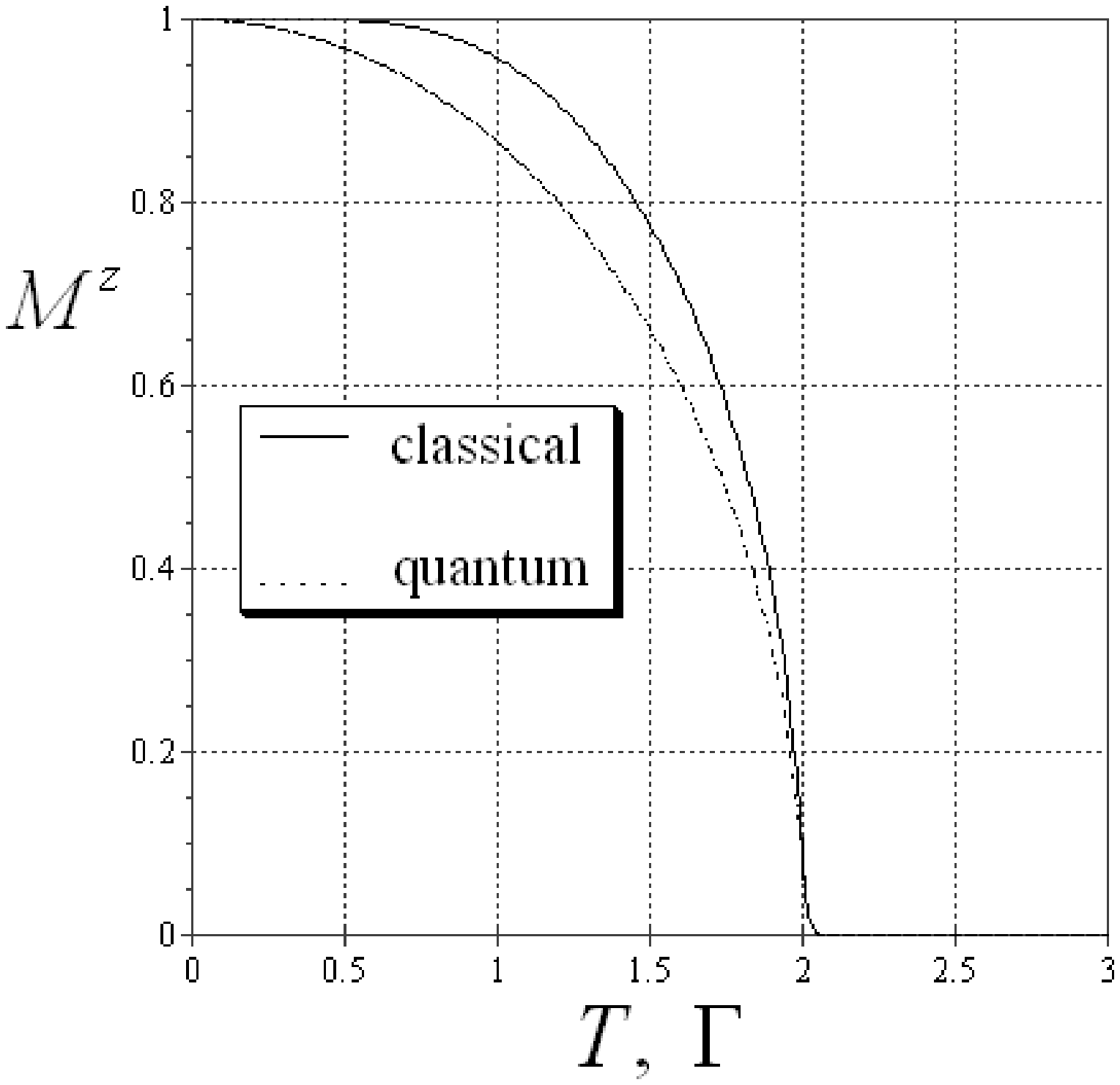}\hspace{-1cm}
\includegraphics[width=8.6cm]{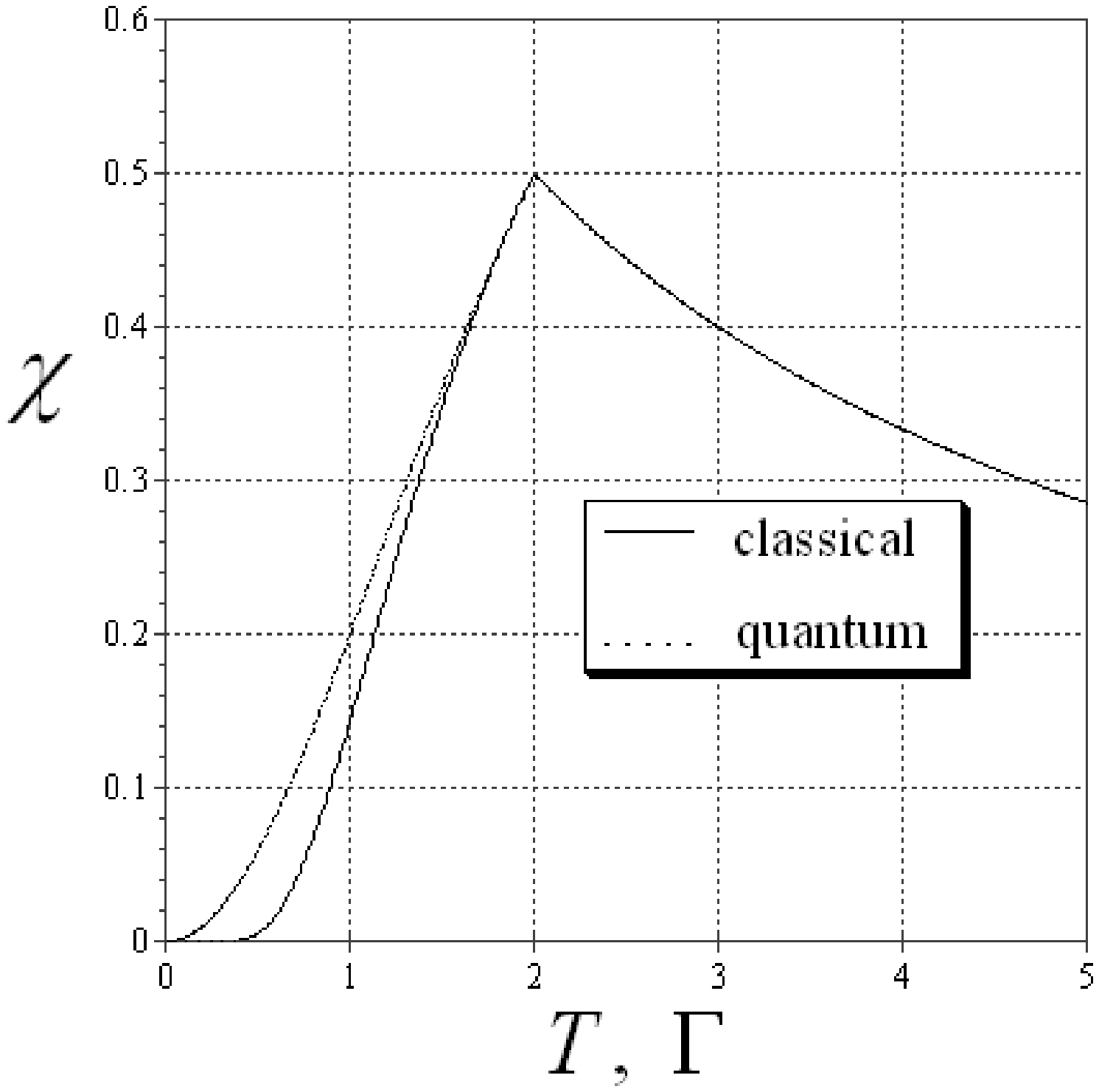}
\end{center}
\caption{\footnotesize 
The longitudinal 
magnetization $M^{z}$ as a function of 
$T$ and $\Gamma$ for 
uniform systems (the left panel). 
The critical points of the second order 
phase transition are 
given by 
$T_{N}=\Gamma_{N}=1/2J$. 
The right panel shows 
the corresponding susceptibility $\chi$ as a 
function of $T$ and $\Gamma$.}
\label{fig:fg1}
\end{figure}

We first consider the case of 
uniform field \cite{Cond}, i.e.,
in (\ref{eq:free_dens}) or 
(\ref{eq:Ma}),
$\hat{h} \to 1$.  
In this limit, 
by using the fact 
$\int_{-\infty}^{\infty}d \hat{h} P(\hat{h})=1$, 
the free energy density $f$ reads
\begin{equation}
f =  
\sum_{l=A,B} f_{l}^{U}; \,\,\,
f_{l}^{U} = 
-J M_{l}^{z} M_{l+1}^{z} 
- \beta^{-1} \log 2 \cosh 
\beta 
\sqrt{
(2JM_{l}^{z}-h)^{2}+
\Gamma_{l+1}^{2}}
\label{eq:Uniform_free}
\end{equation}
and the saddle point equations with respect to 
$M_{A}^{z}$ and $M_{B}^{z}$ are obtained as follows. 
\begin{equation}
M_{l}^{z} =  
\frac{(-2JM_{l+1}^{z}+h)}
{\sqrt{(-2JM_{l+1}^{z}+h)^{2}+\Gamma_{l}^{2}}}
\tanh \beta 
{\sqrt{(-2JM_{l+1}^{z}+h)^{2}+\Gamma_{l}^{2}}}, \,\,\,
(l=A,B). 
\end{equation}
Before we 
investigate the quantum effects, 
we check the classical case,
that is 
$\Gamma_{A}=\Gamma_{B}=0$. 
Then, the above saddle point equations are 
reduced to 
\begin{equation}
M_{l}^{z} =  \tanh \beta (
-2JM_{l+1}^{z}+h), \,\,\,\,
(l=A,B).
\end{equation}
In order to 
determine 
the N$\acute{\rm e}$el temperature, 
we expand the equations 
around 
$M_{A}, M_{B} \simeq 0$ for 
$h=0$ and obtain 
$M_{A}^{z} \simeq 
-2J\beta M_{B}^{z}, 
M_{B}^{z} \simeq 
-2J\beta M_{A}^{z}$. 
From these linearized equations, 
we find that 
only possible solution 
for the case of 
$M_{A}^{z}=M_{B}^{z}=M^{z}$ is 
$M^{z}=0$; whereas with N$\acute{\rm e}$el ordering  
a finite value of 
$M^{z}$ can be 
obtain: 
$M_{A}^{z}=-M_{B}^{z}=-M^{z}$. 
This gives 
the N$\acute{\rm e}$el 
temperature 
$T_{N}=\beta_{N}^{-1}=2J$. 
The linear susceptibilities $\chi_{A}$ and 
$\chi_{B}$ are then
evaluated as 
\begin{equation}
\chi_{l} = 
\lim_{h \to \infty}
\frac{\partial M_{l}^{z}}{\partial h} =  
\frac{\beta (1- 2J\chi_{l+1})}
{\cosh^{2} \beta (-2J M_{l+1}^{z})},\,\,\,
(l=A,B).
\end{equation}
The behavior 
beyond the 
N$\acute{\rm e}$el temperature $T > T_{N}$ is 
determined 
by the condition 
$M_{A}^{z} \simeq 0, 
M_{B}^{z} \simeq 0$, i.e., 
$\chi_{A} = 
\beta (1-2J\chi_{B})$ and 
$\chi_{B}=
\beta (1-2J\chi_{A})$. 
This leads to 
\begin{eqnarray}
\chi & = &  
\frac{2}{T+T_{N}}
\end{eqnarray}
where 
we defined 
$\chi=
\chi_{A}+\chi_{B}$. 
Therefore, 
in the limit of $T \to T_{N}=J$, 
the linear susceptibility 
$\chi$ 
converges to 
$\chi \to 1/2J$. 
On the other hand, 
below the N$\acute{\rm e}$el 
temperature $T < T_{N}$, 
we find the solution for 
the N$\acute{\rm e}$el ordering : 
$M_{A}^{z} = -M_{B}^{z} =-M^{z} \neq 0$ for $h=0$. 
This condition should be satisfied for (cf. Eqn. (31)) 
\begin{eqnarray}
\chi & = & 
\frac{2\beta}{2\beta J + 
\cosh^{2}
\beta (2JM^{z})}
\end{eqnarray}
and at the critical 
point $T=T_{N}, M=0$, 
the susceptibility 
takes $\chi = 2/(2J+T_{N})=1/2J$. 
In Fig. \ref{fig:fg1} (left) 
we plot the shape 
of the susceptibility $\chi$ as 
a function of $T$. 
From 
this figure we find that 
the susceptibility has a cusp, 
instead of the divergence
as observed in the ferromagnetic systems, 
at the 
critical temperature $T=T_{N}$, 
as observed in the analysis for finite 
range models by using mean-field approximations 
\cite{Kittel}. As mentioned earlier, this model being
recastable exactly to a (classical) single-spin in an effective field
(i.e., the mean-field approximation for the partition function 
being exact), there is no scope of any non-trivial wave-function 
with entangled neighbouring spins. 

We next consider the quantum 
case $\Gamma_{A},\Gamma_{B} \neq 0$. 
For simplicity, 
we consider the case of 
the symmetric 
transverse  field, namely, $\Gamma_{A}=\Gamma_{B}=\Gamma$. 
In order to consider the pure quantum 
effects, 
we take 
the limit of $\beta \to \infty$ and 
deal with the following 
coupled equations 
\begin{equation}
M_{l}^{z} =  
\frac{-2JM_{l+1}^{z}+h}
{\sqrt{(-2JM_{l+1}^{z}+h)^{2}+
\Gamma^{2}}},\,\,\,
(l=A,B).
\label{eq:m_az}
\end{equation}
It is important for us to 
bear in mind that 
for $J >0$, 
the above 
equations 
have a solution 
$M^{z}=0$ if $M_{A}=M_{B}$ and 
a solution 
$M^{z} \neq 0$ if 
$M_{A}=-M_{B},$ as expected. 
To determine 
the critical transverse field, 
we expand the above equations 
around 
$M_{A}^{z},M_{B}^{z} \simeq 0$ for $h=0$ as 
$M_{A}^{z} \simeq 
-2JM_{B}^{z}/\Gamma$ and $ M_{B}^{z} \simeq 
-2JM_{A}^{z}/\Gamma$. 
This gives the critical 
point 
$\Gamma_{N}=2J$. 
The spontaneous sub-lattice order $M^z_A$ or $M^z_B$ vanishes
at the N\'eel phase boundary $T_N (\Gamma)$. 
Deep inside the antiferromagnetic phase (at 
$\beta \rightarrow \infty, \Gamma \rightarrow 0, h = 0$), $M_A^z =
1 = -M^z_B$ and the free energy density $f$ can be expressed as $f =  {1}/
{\beta} \log [1 + \exp (-\beta \Delta (\Gamma))]$, the specific heat $\partial^2 f/
\partial T^2 $ will have a variation like $\exp [-\beta \Delta (\Gamma)]$
like that of a two level
system with a gap $\Delta (\Gamma) = {\sqrt {4J^2 + \Gamma^2}}$ here. 
This is the
exact  magnitude of the gap in the magnon spectrum of this long range
transverse Ising antiferromagnet. 

The susceptibilities are 
given by 
\begin{equation}
\chi_{l}  =   
\lim_{h \to 0}
\frac{\partial M_{l}^{z}}
{\partial h} = 
\frac{\Gamma^{2}
(1-2J\chi_{l+1})}
{[(2JM_{l+1}^{z})^{2}+\Gamma^{2}]^{3/2}},\,\,\,\,
(l=A,B)
\end{equation}
Then, 
the behavior 
beyond the critical 
amplitude of 
the transverse field $\Gamma_{N}$ 
is determined 
by the condition 
$M_{A}^{z},M_{B}^{z} \simeq 0$, 
namely, 
$\chi_{A} = 
(1-2J\chi_{B})/\Gamma, \chi_{B} =
(1-2J\chi_{A})/\Gamma$. 
This leads to 
\begin{eqnarray}
\chi & = & \chi_{A}+\chi_{B}=
\frac{2}{\Gamma + \Gamma_{N}}.
\end{eqnarray}
On the other hand, 
below the critical amplitude of the 
transverse field $\Gamma_{N}$, 
we set $M_{A}^{z}=M_{B}^{z}=-M^{z}$ and obtain 
\begin{eqnarray}
\chi & = &  
\frac{2\Gamma^{2}}
{[(2JM)^{2}+\Gamma^{2}]^{3/2}
+
2J\Gamma^{2}}.
\end{eqnarray}
At the critical point $\Gamma_{N}$ ($M^{z}=0$), 
the susceptibility is 
given as 
$\chi=2/(\Gamma_{N}+2J)=1/2J$. 
Around 
$\Gamma \simeq 0$, 
the susceptibility behaves as 
$\chi = 2\Gamma^{2}/(2J)^{3}$. 
In Fig. \ref{fig:fg1} (right), 
we plot $\chi$ as a function of 
$\Gamma$ (and also $T$). 
From this figure, we find that 
the susceptibility has a cusp at the 
critical amplitude of 
the tunneling field. 

\begin{figure}[ht]
\begin{center}
\includegraphics[width=8.6cm]{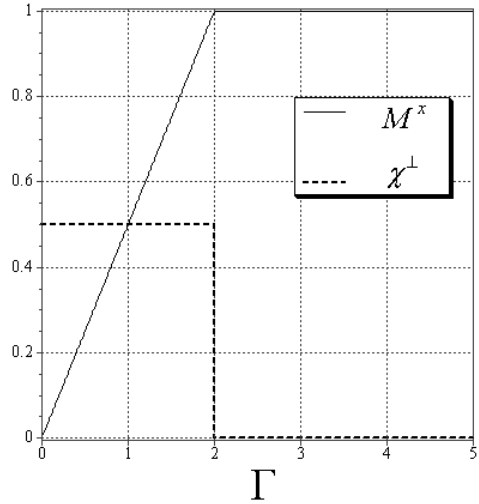}
\end{center}
\caption{\footnotesize 
Longitudinal magnetization $M^{x}$ and 
susceptibility $\chi^{\perp}$.}
\label{fig:fg2}
\end{figure}

We next investigate the 
transverse component of the 
susceptibility. 
The magnetization 
of the transverse direction 
are calculated 
the derivative of the free energy density 
with respect to 
the amplitudes of 
the transverse field 
$\Gamma_{A},\Gamma_{B}$. 
\begin{equation}
M_{l}^{x} = 
\frac{\partial f}{\partial \Gamma_{l}} = 
\frac{\Gamma_{l}}{\sqrt{(-2JM_{l+1}^{z}+h)^{2}+
\Gamma_{l}^{2}}}
\tanh 
\beta \sqrt{(2JM_{l+1}^{z}+h)^{2}+
\Gamma_{l}^{2}},\,\,\,
(l=A,B)
\end{equation}
At the ground state, 
these coupled equations are simplified as 
follows. 
\begin{equation}
M_{l}^{x}  =  
\frac{\Gamma_{l}}
{\sqrt{(-2JM_{l+1}^{z}+h)^{2}+
\Gamma_{l}^{2}}},\,\,\,
(l=A,B)
\end{equation}
In para-magnetic phase is 
specified by $M_{A}^{z}=M_{B}^{z}=0$ 
and 
this gives 
$M_{A}^{x}=M_{B}^{x}=1$. 
On the other hand, 
antiferromagnetic 
phase for 
$h=0$, 
we obtain from (\ref{eq:m_az}) as 
$M_{A}^{z}=
-JM_{B}^{z}/\sqrt{(2JM_{B}^{z})^{2}+
\Gamma_{A}^{2}}, 
M_{B}^{z}=
-JM_{A}^{z}/\sqrt{(2JM_{A}^{z})^{2}+
\Gamma_{B}^{2}}$. 
Hence, 
\begin{equation}
M_{l}^{x}  =  
\frac{\Gamma_{l}}{2J}, \,\,\,
(l=A,B)
\end{equation}
for the antiferromagnetic 
case $M_{A}^{z}=-M_{B}^{z}=-M$. 
Therefore, 
the susceptibilities of 
the transverse direction 
lead to 
\begin{equation}
\chi^{\perp}=\chi_{l}^{x} =  
\frac{\partial M_{l}^{x}}
{\partial \Gamma_{l}} = 0, \,\,\,
(l=A,B)
\end{equation}
for $\Gamma_{A},\Gamma_{B} > \Gamma_{N}$ and 
\begin{equation}
\chi^{\perp} = \chi_{A}^{x} =  \chi_{B}^{x}=\frac{1}{J}
\end{equation}
for $\Gamma_{A},\Gamma_{B} < \Gamma_{N}$. 
We plot the 
transverse 
magnetization $M^{x}=M_{A}^{x}=M_{B}^{x}$ for $M^{z}=M_{A}^{z}=M_{B}^{z}$ in 
Fig. \ref{fig:fg2}. 
We next consider the quantum 
antiferromagnetic system under random fields. 
\section{Analysis under random fields}
In the previous section, 
we investigated the critical phenomena 
for spatially uniform systems. 
It has been conjectured that 
fluctuation in the random (longitudinal) 
field Ising model 
(RFIM) gives rise to a 
many valley structure in the 
configuration space, similar to the case in 
spin glasses. 
The study of the longitudinal 
random field transverse field Ising model 
with ferromagnetic uniform interactions has 
already been made \cite{Dutta}.
There seems to be no reported analytic 
research for the RFIM with antiferromagnetic 
interactions in transverse field. 
It should be interesting to investigate the 
competition 
between two different kinds of 
frustration; 
frustration due to the quenched disorder and 
the frustration induced by the antiferromagnetic 
interactions. In this section, we investigate 
quantum systems under random on-site
longitudinal fields.
In this paper, 
we consider the following two cases of 
the random field distributions : 
\begin{eqnarray}
P_{g}(\mbox{\boldmath $h$}) & = &
\prod_{i=1}^{N}
\frac
{1}{\sqrt{2\pi}\sigma}
{\exp}
\left[
-\frac{1}{2\sigma^{2}}
(h_{i}-h_{0})^{2}
\right] \equiv 
\prod_{i=1}^{N}
P_{g}(h_{i}) 
\label{eq:gauss} \\
\rm{and}\quad P_{b}(\mbox{\boldmath $h$}) & = & 
\prod_{i=1}^{N}
\left\{
\theta \delta (h_{i}-h_{0}) + 
(1-\theta) 
\delta (h_{i}+h_{0})
\right\} \equiv 
\prod_{i=1}^{N}
P_{b}(h_{i}),
\label{eq:binary}
\end{eqnarray}
where the bias factor of the 
binary random field $\theta$ takes 
$0 < \theta <1$. 
For these distributions, 
the free energy densities 
$f_{g,b} \equiv (1/N)\int_{-\infty}^{\infty}
d\hat{h} P_{g,b}(\hat{h}) F(\hat{h})$ 
become
\begin{eqnarray}
f_{g} & = & 
\sum_{l=A,B}
f_{l}^{g} \\
f_{l}^{g} & = & 
-J M_{l+1}^{z}
M_{l}^{z} - 
\beta^{-1}
\int_{-\infty}^{\infty}
Dx 
\log 2 \cosh \beta
\sqrt{
\{-2JM_{l+1}^{z}+h(\sigma x +h_{0})\}^{2}+
\Gamma_{l}^{2}}, 
\end{eqnarray}
for the Gaussian random field 
$P_{g}(\mbox{\boldmath $h$})$, where
$Dx \equiv \frac{1}{\sqrt{2\pi}}e^{{-x^{2}/2}}dx$
and 
\begin{eqnarray}
f_{b} & = & 
\sum_{l=A,B} 
f_{l}^{b} \\
f_{l}^{b} & = & 
-J M_{l+1}^{z} M_{l}^{z} 
-  
\beta^{-1} \theta 
\log 2 
\cosh \beta 
\sqrt{
\{-2M_{l+1}^{z}J +
hh_{0}\}^{2}+
\Gamma_{l}^{2}} \nonumber \\
\mbox{} & - & \beta^{-1}(1-\theta)
\log 2 
\cosh \beta 
\sqrt{
\{2M_{l+1}^{z}J +hh_{0}\}^{2}+
\Gamma_{l}^{2}}, 
\end{eqnarray}
for the binary random field 
$P_{b}(\mbox{\boldmath $h$})$. 
In following, 
we investigate the critical phenomena 
given by these free energy densities 
for these two cases. 
\subsection{The Gaussian random field}
\begin{figure}[ht]
\begin{center}
\includegraphics[width=8.6cm]{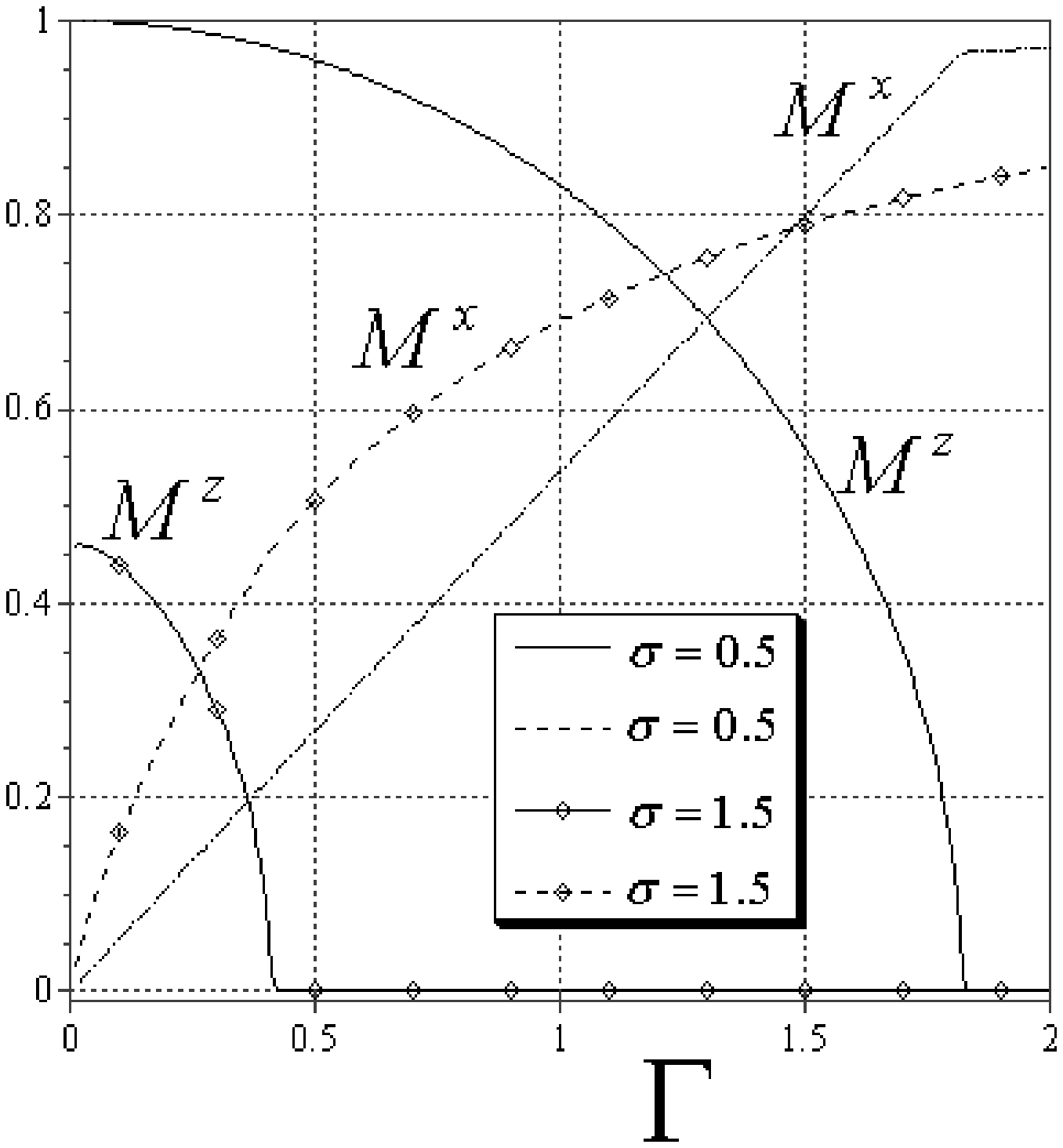}\hspace{-1cm}
\includegraphics[width=8.6cm]{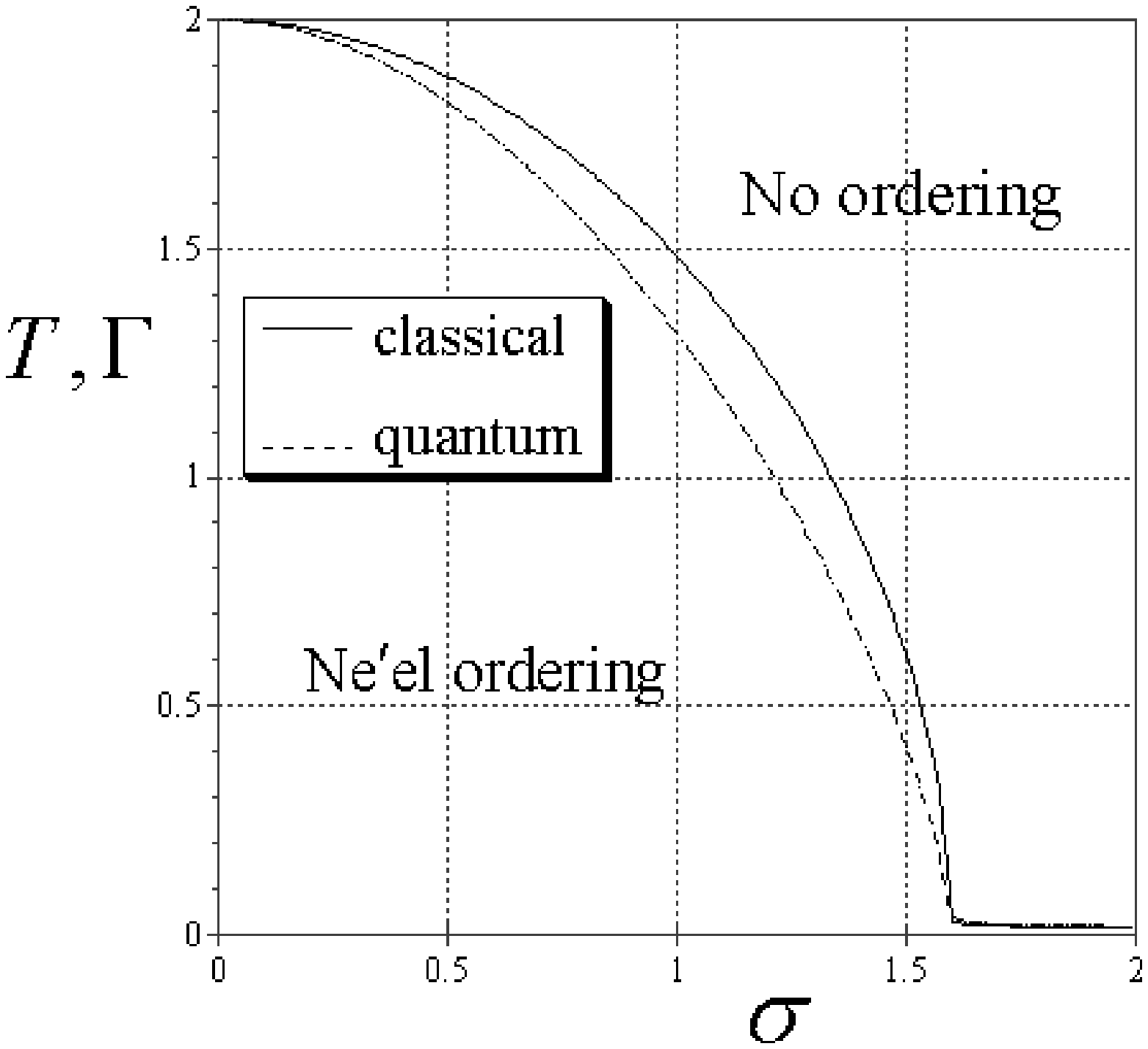}
\end{center}
\caption{\footnotesize 
The right panel 
shows 
phase boundaries 
between 
antiferromagnetic and 
paramagnetic 
phases for classical and 
quantum systems for $(J,h,h_{0})=(1,1,1)$. 
The left panel 
shows the variation of longitudinal and transverse 
magnetizations 
$M^{z}$ and $M^{x}$ with 
$\Gamma$.
}
\label{fig:fg3}
\end{figure}

\begin{figure}[ht]
\begin{center}
\includegraphics[width=8.5cm]{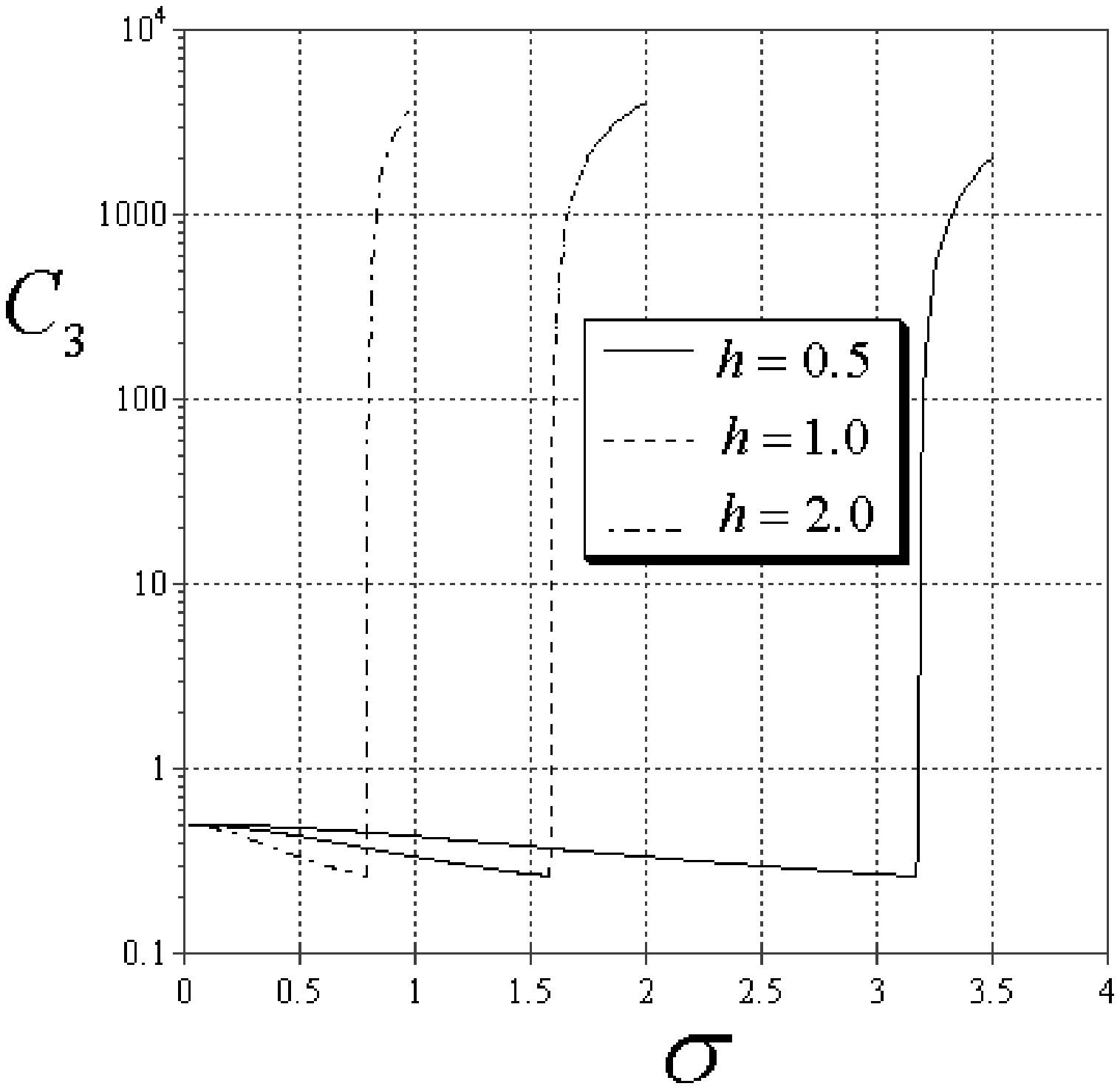}\hspace{-0.8cm}
\includegraphics[width=8.5cm]{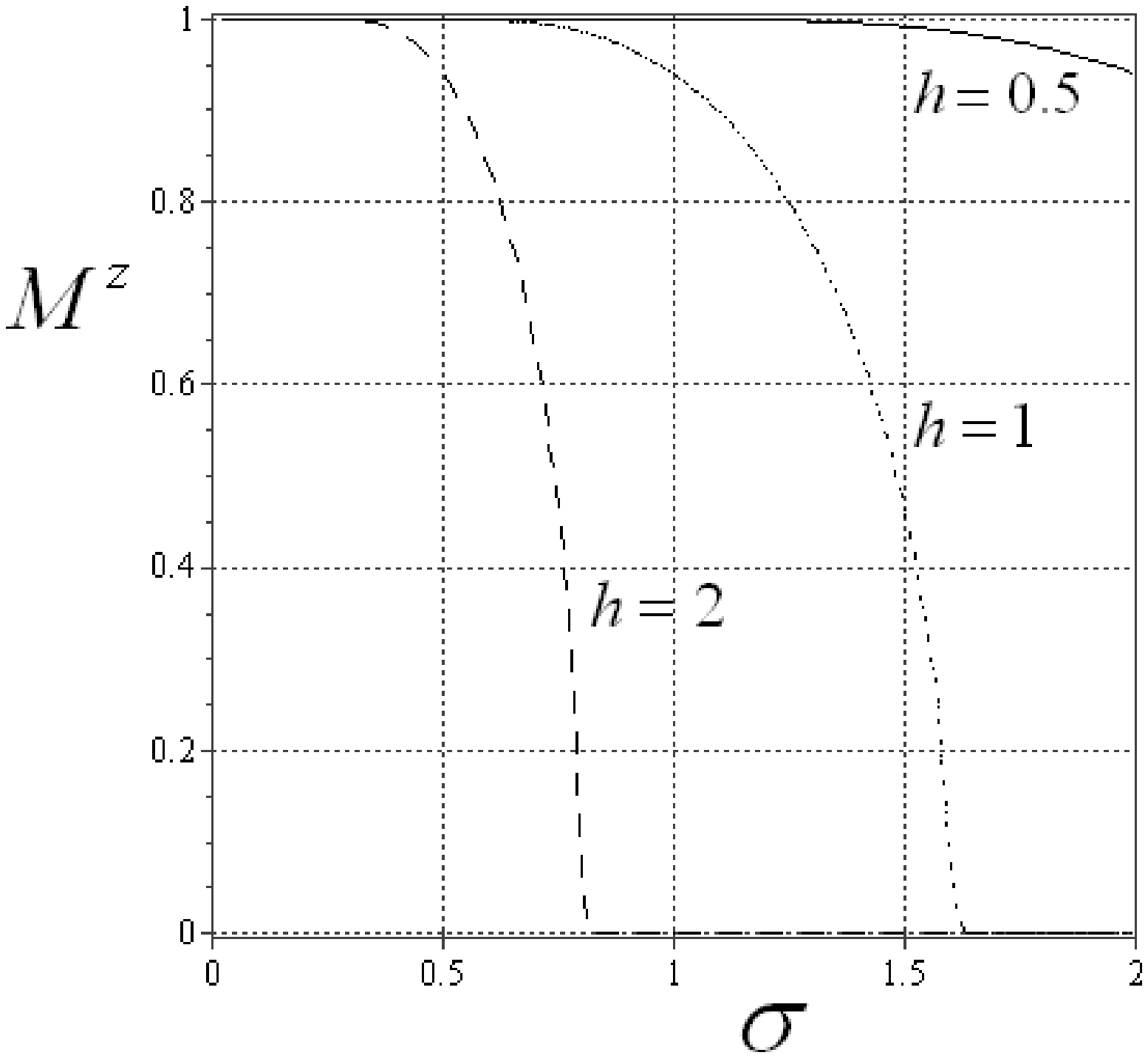}
\end{center}
\caption{\footnotesize 
The longitudinal 
magnetization $M^{z}$ 
as a function of 
$\sigma$ for several values of 
$h_{0}$ for the case of $\Gamma=0$ (right panel). 
Each line of 
a solution of the equation (\ref{eq:mGamma0}). 
The left panel 
shows the $\sigma$-dependence of the 
factor $C_{3}$ of 
the third order 
of the magnetization near the critical point.}
\label{fig:fg4}
\end{figure}

For the Gaussian random field (\ref{eq:gauss}) 
the saddle point equations 
are given by the derivative of 
the free energy density $f_{G}$ with 
respect to $M_{A}^{z}$ and 
$M_{B}^{z}$. 
Then we have  
\begin{equation}
M_{l}^{z} =   
\int_{-\infty}^{\infty}
Dx 
\frac{\{
-2JM_{l+1}^{z}+h(\sigma x + h_{0})\}}
{\sqrt{\{
-2JM_{l+1}^{z}+h(\sigma x + h_{0})\}^{2}+
\Gamma_{l}^{2}}}
\tanh \beta 
\sqrt{\{
-2JM_{l+1}^{z}+h(\sigma x + h_{0})\}^{2}+
\Gamma_{l}^{2}}
\end{equation}
for $l=A,B$. 
At the ground state 
$(\beta \to \infty)$ 
these equations of states are 
simplified as follows 
\begin{equation}
M_{l}^{z}  =   
\int_{-\infty}^{\infty}
Dx
\frac{\{-2JM_{l+1}^{z}+h(\sigma x + h_{0})\}}
{\sqrt{
\{-2JM_{l+1}^{z}+h(\sigma x + h_{0})\}^{2}
+
\Gamma_{l}^{2}
}},\,\,\,
(l=A,B).
\label{eq:m_az2}
\end{equation}
For the possible choice 
$M_{A}^{z}=M^{z}=-M_{B}^{z}$ to 
detect the N$\acute{\rm e}$el 
ordering, we solve the above equations 
for the symmetric 
transverse fields 
$\Gamma_{A}=\Gamma_{B}=\Gamma$ numerically. 
In Fig. \ref{fig:fg3} (left), 
we plot the case of 
the center of the Gaussian, 
$h_{0}$ takes 
$h_{0}=0$ and 
the deviation of the Gaussian 
$\sigma=0.5$ and $1.5$. 
From this figure, 
we find that 
the system undergoes second-order phase transition 
at the critical amplitude of 
the transverse field 
from the behavior of $M^{z}$. 
If the phase transition is 
first order for the case of the center $h_{0}$ of 
the Gaussian (\ref{eq:gauss}) is zero, 
we can expand the saddle point equation 
with respect to $M_{A}^{z}$ under the condition 
$\Gamma_{A}=\Gamma_{B}=\Gamma$ and 
$M_{B}^{z}=-M_{A}^{z}=-M^{z}$ as 
\begin{eqnarray}
M^{z} & = &  
C_{1} M^{z}-C_{3}(M^{z})^{3}+ 
{\cal O}((M^{z})^{5})
\end{eqnarray}
with 
\begin{equation}
C_{1} =  2J\Gamma^{2}
\int_{-\infty}^{\infty}
\frac{Dx}
{[(h\sigma x)^{2}+\Gamma^{2}]^{3/2}},\,\,\,\,\,
C_{3} =  
4J^{3}\Gamma^{2}
\int_{-\infty}^{\infty}
Dx\, 
\frac{2(h\sigma x)^{2}+\Gamma^{2}}
{[
(h\sigma x)^{2}+\Gamma^{2}]^{7/2}}.
\end{equation}
The phase boundary $\Gamma (\sigma)$ of the 
continuous transition 
between the N$\acute{\rm e}$el and 
the paramagnetic phases is 
obtained for a given set of 
the parameters $(J,h_{0},h)$ 
by the condition $a=1$, namely
\begin{equation}
\Gamma = 
\left(
2J
\int_{-\infty}^{\infty}
\frac{Dx}
{[(h\sigma x)^{2}+\Gamma^{2}]^{3/2}}
\right)^{-1/2}
\label{eq:boundary}.
\end{equation}
The second order phase transition 
is observed 
for $C_{3} >0$, whereas
a first order phase transition 
is found for $C_{3} < 0$. 
In Fig. \ref{fig:fg3} (right), 
we plot the boundaries 
between the N$\acute{\rm e}$el and 
the paramagnetic phases 
for both quantum and classical systems.  
In this plot, 
we set $(J,h,h_{0})=(1,1,0)$. 
In the left panel of Fig. \ref{fig:fg4}, 
we plot the factor $C_{3}$ of the third order 
of the expansion of the magnetization $M^{z}$ 
as a function of $\sigma$. 
In this plot we 
substituted 
the solution of 
the boundary (\ref{eq:boundary}) for a given $\sigma$ 
into $C_{3}$. 
From this panel 
we find, 
that from the value 
$C_{3}=4(J/\Gamma)^{3} >0$ at $\sigma=0$, that
$C_{3}$ decreases and 
takes its positive minima at just below 
the critical point $\sigma_{c}$, and 
beyond the critical point 
$C_{3}$ increases again. We 
therefore conclude that 
the phase transition is always second order. 
We might see this from the argument below. 
In the limit of $\Gamma \to 0$, 
the equation of state 
(\ref{eq:m_az2}) for 
$M^{B}=-M_{A}^{z}=-M^{z}$ 
is simplified to 
\begin{eqnarray}
M^{z} & = & 
\int_{-\infty}^{\infty}
Dx\,
{\rm sgn}
\left(
2JM^{z}+h\sigma x
\right) = 
1-2H
\left(
\frac{2JM^{z}}
{h\sigma}
\right)
\label{eq:mGamma0}.
\end{eqnarray}
In Fig. \ref{fig:fg4} (right panel), 
we plot the solution of (\ref{eq:mGamma0}) 
for several values of 
$h$. 
We found that 
for $\sigma \to \infty$, 
$H(2JM^{z}/h\sigma) \to 1/2$ and 
$M^{z} \to 0$, 
whereas, 
for $\sigma \to 0$, 
$H(2JM^{z}/h\sigma) \to 0$ and 
$M^{z} \to 1$.
By expanding (\ref{eq:mGamma0}) with respect to 
$M^{z}$ up to the first order, 
we obtain the critical point $\sigma_{c}$ as 
\begin{equation}
\sigma_{c} = 
\sqrt{\frac{2}{\pi}}
\left(
\frac{2J}{h}
\right).
\end{equation}
The magnetization varies continuously
near this critical point (of the second order phase transition): 
$M^{z}=\sqrt{12\sigma/\pi \sigma_{c}}
(1-\sigma/\sigma_{c})^{1/2}$. 

We next evaluate 
the $x$-component 
(the transverse component) 
of the magnetizations 
$M_{A}^{x}$ and 
$M_{B}^{x}$. 
These are calculated at the ground state 
$\beta \to \infty$ as  
\begin{equation}
M_{l}^{x} = 
\frac{\partial f_{G}}{\partial \Gamma_{l}}=
\int_{-\infty}^{\infty}
\frac{\Gamma_{l} Dx}
{\sqrt{
\{-2JM_{l+1}^{z}+h(\sigma x + h_{0})\}^{2}
+\Gamma_{l}^{2}}},\,\,\,
(l=A,B)
\end{equation}
In Fig. \ref{fig:fg3} (left panel), 
we plot the results. 
It should be noted that 
in the limit of 
$\Gamma \to \infty$, 
$M^{x}$ saturates 
$M^{x} \simeq \int_{-\infty}^{\infty}
Dx (\Gamma/\Gamma)=1$.
\subsection{The binary random fields}
\begin{figure}[ht]
\begin{center}
\includegraphics[width=8.6cm]{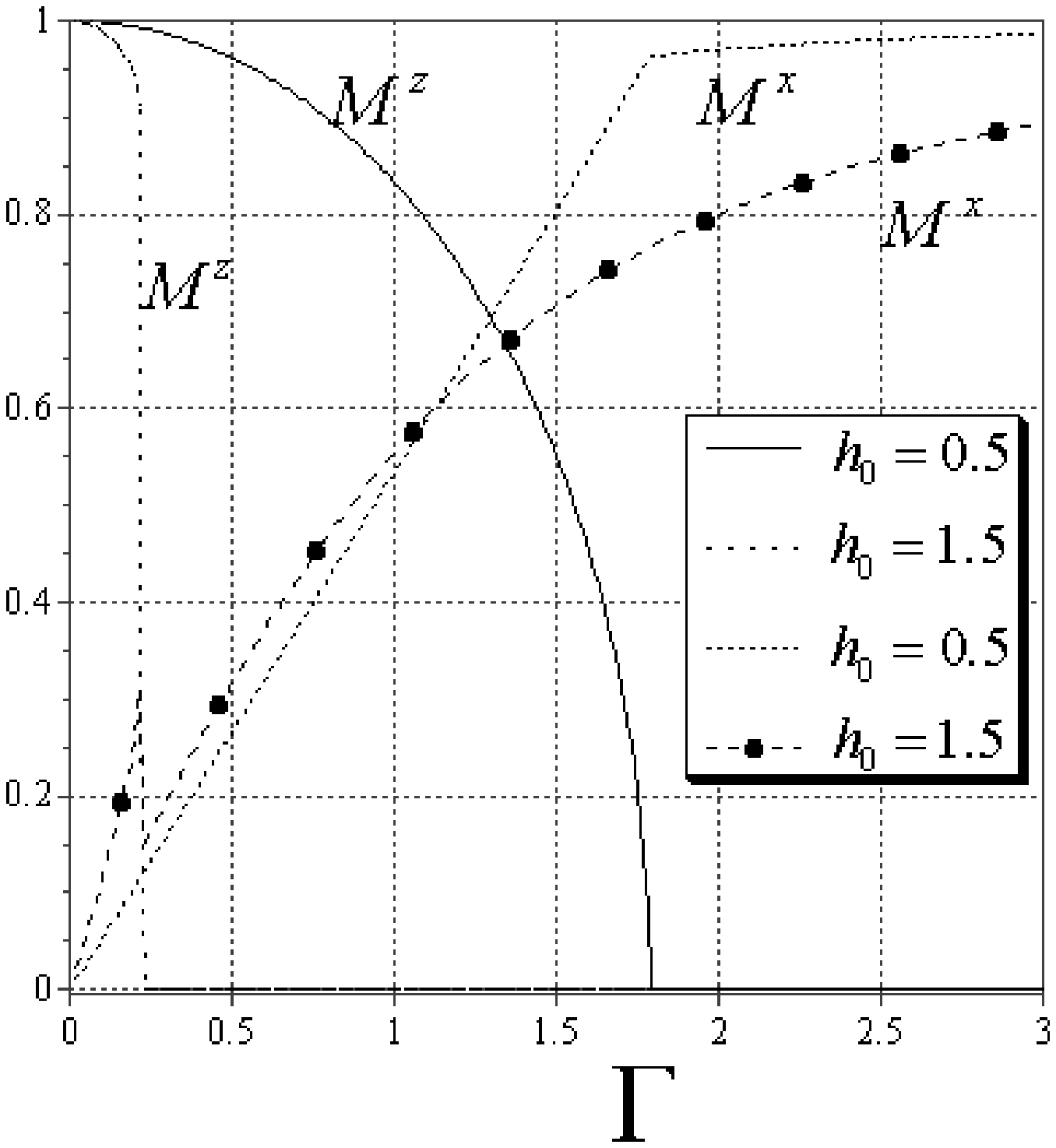}\hspace{-1cm}
\includegraphics[width=8.6cm]{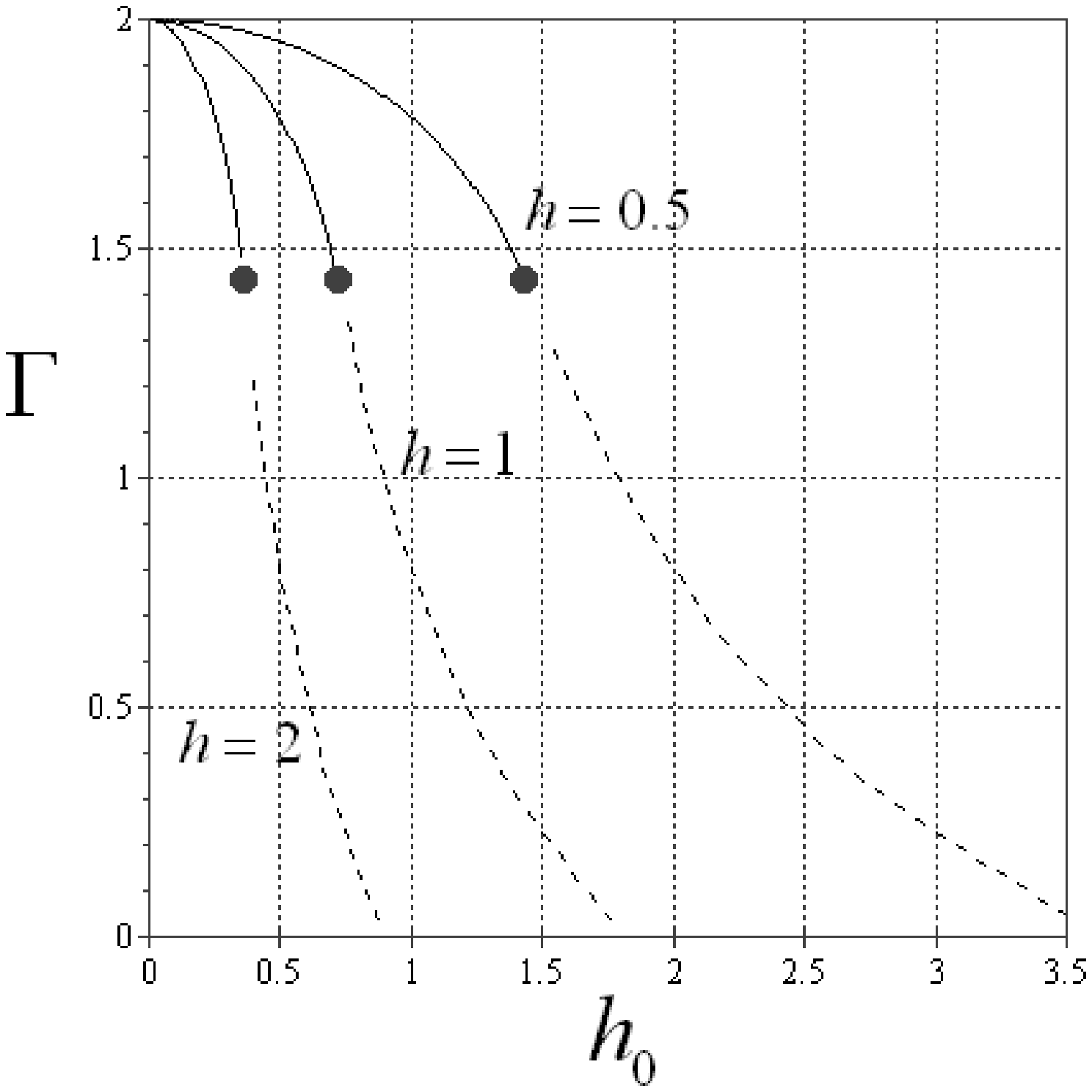}
\end{center}
\caption{\footnotesize 
The longitudinal and 
the transverse magnetizations, 
$M^{z}$ and $M^{x}$ 
as a function of 
$\Gamma$ 
for the case of the binary random fields 
(right panel). 
The left panel shows 
phase boundaries 
between 
antiferromagnetic and 
paramagnetic 
phases. The dots represent 
tri-critical points
$(h_{0}^{c},\Gamma_{c})
=((1/h)(4/5)^{3/2}, 2(4/5)^{3/2})$ 
we set $J=1$. 
}
\label{fig:fg5}
\end{figure}

For the binary random 
fields, we obtain 
the saddle point equations 
by taking the derivative of the free 
energy density $f_{B}$ 
with respect to $M_{A}^{z}$ and 
$M_{B}^{z}$ as follows: 
\begin{eqnarray}
M_{l}^{z} & = & 
-
\frac{\theta (2JM_{l+1}^{z}-hh_{0})}
{\sqrt{(2JM_{l+1}^{z}-hh_{0})^{2}+
\Gamma_{l}^{2}}}
\tanh \beta 
\sqrt{(2JM_{l+1}^{z}-hh_{0})^{2}+
\Gamma_{l}^{2}} \nonumber \\
\mbox{} & - & 
\frac{(1-\theta)(2JM_{l+1}^{z}+hh_{0})}
{\sqrt{(2JM_{l+1}^{z}+hh_{0})^{2}+
\Gamma_{l}^{2}}}
\tanh \beta 
\sqrt{(2JM_{l+1}^{z}+hh_{0})^{2}+
\Gamma_{l}^{2}},\,\,\,
(l=A,B).
\end{eqnarray}
At the ground state $\beta \to \infty$, 
these equations are 
simplified to 
\begin{equation}
M_{l}^{z} =  
-\frac{\theta (2JM_{l+1}^{z}-hh_{0})}
{\sqrt{(2JM_{l+1}^{z}-hh_{0})^{2}+
\Gamma_{l}^{2}}} - 
\frac{(1-\theta) (2JM_{l+1}^{z}+hh_{0})}
{\sqrt{(2JM_{l+1}^{z}+hh_{0})^{2}+
\Gamma_{l}^{2}}},\,\,\,
(l=A,B). 
\label{eq:m_za3}
\end{equation}
We solve the above 
equations 
numerically and 
plot it in Fig. \ref{fig:fg5} (left panel). 
From 
this figure, we find that 
the system undergoes 
first order 
phase transition 
when the value of 
$h_{0}$ is larger than 
same critical point $h_{0}^{c}$. 
Whereas, for small value of 
$h_{0}<h_{0}^{c}$, the phase transition is 
the second order. 

In following, 
we determine the tri-critical point 
$(h_{0}^{c},\Gamma_{c})$. 
If the transition is continuous, 
we can expand the saddle point equation for 
$M_{A}^{z}$ under the condition 
$\Gamma_{A}=\Gamma_{B}=\Gamma$ and 
$M_{B}^{z}=-M_{A}^{z}=-M^{z}$ as follows: 
\begin{eqnarray}
M^{z} & = & 
\tilde C_{0} 
+\tilde C_{1} M^{z}
+\tilde C_{2} (M^{z})^{2} + \tilde C_{3} (M^{z})^{3}+ 
{\cal O}((M^{z})^{4})
\end{eqnarray}
where we defined 
\begin{eqnarray}
\tilde C_{0} & = & 
\frac{2(\theta -1)hh_{0}}{\sqrt{(hh_{0})^{2}+\Gamma^{2}}} \\
\tilde C_{1} & = & 
\frac{2J\Gamma^{2}}{\{(hh_{0})^{2}
+\Gamma^{2}\}^{3/2}} \\
\tilde C_{2} & = & 
-2J^{2} hh_{0}
(2\theta-1)
\left[
\frac{\Gamma^{2}+4(hh_{0})^{2}}
{
\{(hh_{0})^{2}+
\Gamma^{2}
\}^{5/2}
}
\right] \\
\tilde C_{3} & = & 
4J^{3}
\left[
\frac{\Gamma^{4}-4\Gamma^{2}
(hh_{0})^{2}}
{
\{(hh_{0})^{2}+\Gamma^{2}
\}^{7/2}
}
\right]. 
\end{eqnarray}
Therefore, 
if the distribution 
(\ref{eq:binary}) is symmetric 
 (i.e., $\theta = 1/2$), 
the factors 
$\tilde C_{0}$ and $\tilde C_{2}$ vanish and 
the magnetization behaves as 
\begin{eqnarray}
M^{z} & = & \tilde C_{1} M^{z} + 
\tilde C_{3} (M^{z})^{3} + 
{\cal O}((M^{z})^{5}).
\end{eqnarray}
From this expression, 
we find that 
a second order phase transition is 
found when the condition 
$\tilde C_{1}=1$ and $\tilde C_{3} < 0$ holds. 
On the other hand, 
a first order phase transition is 
observed for $\tilde C_{1}=1$ and 
$\tilde C_{3} > 0$. 
Therefore, 
the point $(h_{0}^{c},\Gamma_{c})
=((J/h)(4/5)^{3/2},2J(4/5)^{3/2})$,
which is determine by 
$\tilde C_{1}=1,\tilde C_{3}=0,$  
corresponds to 
a tri-critical point on the phase boundary. 
In Fig. \ref{fig:fg5} (right panel) we plot these phase boundaries. 
We should notice that the 
critical point $\Gamma_{c}$ 
is independent of $h$. 
We find 
that for $h_{0} > h_{0}^{c}$, 
the transition from 
the symmetry breaking 
phase to 
symmetric phase is first order. 
To compare 
this result with the case of 
the Gaussian 
random field, 
we consider the limit 
$\Gamma \to 0$ in 
(\ref{eq:m_za3}). 
We obtain  
\begin{equation}
M_{l}^{z} = \theta \, {\rm sgn}
(2JM_{l+1}^{z}+hh_{0})
+
(1-\theta)\, {\rm sgn}(2JM_{l+1}^{z}-hh_{0}), \qquad (l = A,B).
\end{equation}
To detect the transition 
point between 
the N$\acute{\rm e}$el 
and paramagnetic phases, 
we set $M_{l}^{z}=M^{z}=-M_{l+1}^{z}$ and 
$\theta =1/2$ for simplicity. 
We then have 
\begin{equation}
2M^{z} = {\rm sgn}
(2JM^{z}+hh_{0})
+{\rm sgn}(2JM^{z}-hh_{0}).
\end{equation}
Apparently, 
$M_{z}$ takes values $1$ or $0$ and 
the critical point of the first order phase transition 
is determined by $2J-hh_{0}=0$, 
i.e., 
$h_{0}^{c}=2J/h$. 
This point $h_{0}^{c}$ is 
observed on the crossing point 
on the $h_{0}$-axis in 
Fig. \ref{fig:fg5} (right panel). 
On the other hand, 
as we saw in Fig. \ref{fig:fg4} (left panel), 
the magnetization 
$M^{z}$ for the 
Gaussian random field drops gradually and 
the phase transition is second order even if 
there is no quantum fluctuation 
$\Gamma=0$ at the ground state (at $\beta \to \infty$). 
This is a reason why 
the order-disorder phase transition in the Gaussian 
random field Ising model is always first order and 
it does not depend on 
$\Gamma$ or $\sigma$.
Finally, we calculate the transverse magnetization 
$M_{A}^{x}$ and $M_{B}^{x}$. 
From the derivative 
of the free energy density 
$f_{B}$ with respect to $\Gamma_{A}$ and $\Gamma_{B}$, 
we obtain 
\begin{eqnarray}
M_{l}^{x} & = &  
\frac{\partial f_{b}}{\partial \Gamma_{l}} = 
\frac{\theta \Gamma_{l}}
{\sqrt{\{-2JM_{l+1}^{z} + hh_{0}\}^{2}+\Gamma_{l+1}^{2}}}
\tanh \beta 
\sqrt{
\{-2JM_{l+1}^{z}+hh_{0}\}^{2}
+\Gamma_{l}^{2}} \nonumber \\
\mbox{} & + &  
\frac{(1-\theta) \Gamma_{l}}
{\sqrt{\{2JM_{l+1}^{z} + hh_{0}\}^{2}+\Gamma_{l}^{2}}}
\tanh \beta 
\sqrt{
\{2JM_{l+1}^{z}+hh_{0}\}^{2}
+\Gamma_{l}^{2}},\,\,\,
(l=A,B). 
\end{eqnarray}
At the ground state, 
these equations are simplified as 
\begin{equation}
M_{l}^{x}  =   
\frac{\theta \Gamma_{l}}
{\sqrt{\{-2JM_{l+1}^{z} + hh_{0}\}^{2}+\Gamma_{l}^{2}}} +
\frac{(1-\theta) \Gamma_{l}}
{\sqrt{\{2JM_{l+1}^{z} + hh_{0}\}^{2}+\Gamma_{l}^{2}}},\,\,\,
(l=A,B). 
\label{eq:mx_binary}
\end{equation}
In Fig. \ref{fig:fg5} (left panel)
we plot the transverse 
magnetization for case of the 
symmetric amplitude of 
the tunneling field $\Gamma_{A}=\Gamma_{B}=\Gamma$ 
as a function of $\Gamma$. 
Obviously, 
for large $\Gamma$, 
we found 
$M^{x}=1$ from (\ref{eq:mx_binary}). 

As may be noted, in the case where 
the random quenched fields are 
symmetrically distributed about its zero value, the
effective sub-lattice symmetry could be utilized to reduce the whole
problem to that of a ferromagnet.

\section{Summary}
In this paper, 
we proposed 
an analytically solvable 
quantum antiferromagnetic Ising model. Because of
Ising anisotropy and long-range interactions, it has
solvable N\'{e}el-like ground and other state properties.
In view of the extensive recent studies in quantum 
antiferromagnets (\cite{Kaiser}-\cite{Moessner2}),
particularly in quantum Ising antiferromagnets 
(\cite{Moessner},\cite{Moessner2}), this kind of 
analysis should be of considerable importance.

In the analysis of 
spatially uniform system, 
we found the N\'{e}el 
order below  
the tunneling field $\Gamma_{N}$ and 
show that the linear susceptibility 
has a cusp variation around
that critical $\Gamma_{N}$. 
It may be mentioned that a similar behavior in the
half-filled Hubberd 
model was observed earlier \cite{Stefan}.


For this uniform spin system, 
the free energy density (\ref{eq:Uniform_free}) gives the ground state
energy in the zero temperature limit and it also gives the low
temperature behavior of the specific heat, the exponential variation
of which gives the precise gap magnitude 
$\Delta (= {\sqrt{ 4J^2 + \Gamma^2}})$ in the excitation spectrum of
the system. It may be noted that, because of the
restricted (Ising) symmetry and the infinite dimensionality (long
range interaction) involved, there need not be any  conflict
with the Haldane conjecture. Although our entire analysis has been for
spin-1/2 (Ising) case, because of the reduction of the effective
Hamiltonian (\ref{eq:Single_Ham}) to that of a single spin in an effective vector field, the
results can be easily generalized for higher values of the spin $S$. No
qualitative change is observed. 
 The order-disorder transition in the model can be driven both by
thermal fluctuations (increasing $T$) or by the quantum fluctuations
(increasing $\Gamma$). This transition in the model has been investigated 
studying the  behaviors of the (random sub-lattice) staggered
magnetization and the (longitudinal and transverse) susceptibilities.
No quantum phase transition, where the gap $\Delta$ vanishes, is observed
in the model, unlike in the one dimensional transverse Ising antiferromagnets.  

By analysis of 
the disordered 
system as the random field Ising model in a transverse field, 
we found that the order of the phase transition 
changes at a tri-critical point. 
These conditions are obtained 
analytically for both the Gaussian 
random fields and 
the binary random fields. 
We believe, analysis of such model systems 
might provide some insights also for
the quantum antiferromagnets
with short range interactions.

\begin{acknowledgments}
The present work was financially 
supported by 
{\it IKETANI SCIENCE AND 
TECHNOLOGY FOUNDATION} grant no. 
0174004 D. 
One of the authors (B. K. C.) 
thanks Hokkaido University and 
University of Tokyo 
for hospitality.
\end{acknowledgments}

\end{document}